\documentclass[11pt]{article}

\usepackage[T1]{fontenc}
\usepackage[utf8]{inputenc}
\usepackage[margin=1in]{geometry}
\usepackage{times}
\usepackage{microtype}
\usepackage{graphicx}
\usepackage{amsmath,amssymb}
\usepackage[numbers]{natbib}
\usepackage{xcolor}
\usepackage{caption}
\usepackage{booktabs}
\usepackage{url}
\usepackage{hyperref}
\usepackage{float}      % for [H] placement specifier
\usepackage{tabularx}
\usepackage{array}
\usepackage{listings}

\floatstyle{ruled}
\newfloat{algorithm}{tbp}{loa}
\floatname{algorithm}{Algorithm}

\hypersetup{
  colorlinks=true,
  linkcolor=blue!45!black,
  citecolor=blue!45!black,
  urlcolor=blue!45!black
}

\lstdefinelanguage{pseudocode}{
  morekeywords={Input,Output,State,Precondition,function,for,each,in,while,if,else,return,assert,true,false,not,empty},
  sensitive=true,
  morecomment=[l]{\#},
  morestring=[b]"
}

\title{Generative AI for Video Translation: A Scalable Architecture for Multilingual Video Conferencing}
\author{%
Amirkia Rafiei Oskooei\thanks{Corresponding author: \texttt{amirkia.oskooei@std.yildiz.edu.tr}} \and
Eren Caglar \and
Ibrahim \c{S}ahin \and
Ayse Kayabay \and
Mehmet S.\ Aktas\\[4pt]
\small Department of Computer Engineering, Yildiz Technical University, Istanbul 34220, Turkey
}
\date{}

\begin{document}
\maketitle

\begin{abstract}
The real-time deployment of cascaded generative AI pipelines for applications like video translation is constrained by significant system-level challenges. These include the cumulative latency of sequential model inference and the quadratic ($\mathcal{O}(N^2)$) computational complexity that renders multi-user video conferencing applications unscalable. This paper proposes and evaluates a practical system-level framework designed to mitigate these critical bottlenecks. The proposed architecture incorporates a turn-taking mechanism to reduce computational complexity from quadratic to linear in multi-user scenarios, and a segmented processing protocol to manage inference latency for a perceptually real-time experience. We implement a proof-of-concept pipeline and conduct a rigorous performance analysis across a multi-tiered hardware setup, including commodity (NVIDIA RTX 4060), cloud (NVIDIA T4), and enterprise (NVIDIA A100) GPUs. Our objective evaluation demonstrates that the system achieves real-time throughput ($\tau < 1.0$) on modern hardware. A subjective user study further validates the approach, showing that a predictable, initial processing delay is highly acceptable to users in exchange for a smooth, uninterrupted playback experience. The work presents a validated, end-to-end system design that offers a practical roadmap for deploying scalable, real-time generative AI applications in multilingual communication platforms.
\end{abstract}


\medskip
\noindent\textbf{Note:} This manuscript is the authors' accepted version of a paper published in \emph{Applied Sciences} (MDPI), 2025. The final version is available from the publisher at \url{https://www.mdpi.com/2076-3417/15/23/12691}.
%%%%%%%%%%%%%%%%%%%%%%%%%%%%%%%%%%%%%%%%%%
% \setcounter{section}{-1} %% Remove this when starting to work on the template.
% \section{How to Use this Template}

% The template details the sections that can be used in a manuscript. Note that the order and names of article sections may differ from the requirements of the journal (e.g., the positioning of the Materials and Methods section). Please check the instructions on the authors' page of the journal to verify the correct order and names. For any questions, please contact the editorial office of the journal or support@mdpi.com. For LaTeX-related questions please contact latex@mdpi.com.%\endnote{This is an endnote.} % To use endnotes, please un-comment \printendnotes below (before References). Only journal Laws uses \footnote.

% % The order of the section titles is different for some journals. Please refer to the "Instructions for Authors” on the journal homepage.

%%%%%%%%%%%%%%%%%%%%%%%%%%%%%%%%%%%%%%%%%%%%%%%%%%%%%%%%%%%%%%%%%%%%%%%%%%%%%%%%%%%%
%%%%%%%%%%%%%%%%%%%%%%%%%%%%%%%%%%%%%%%%%%%%%%%%%%%%%%%%%%%%%%%%%%%%%%%%%%%%%%%%%%%%
%%%%%%%%%%%%%%%%%%%%%%%%%%%%%%%%%%%%%%%%%%%%%%%%%%%%%%%%%%%%%%%%%%%%%%%%%%%%%%%%%%%%
\section{Introduction}\label{sec:intro}
\begin{figure}[H]
  %  \centering
    \includegraphics[width=1\linewidth]{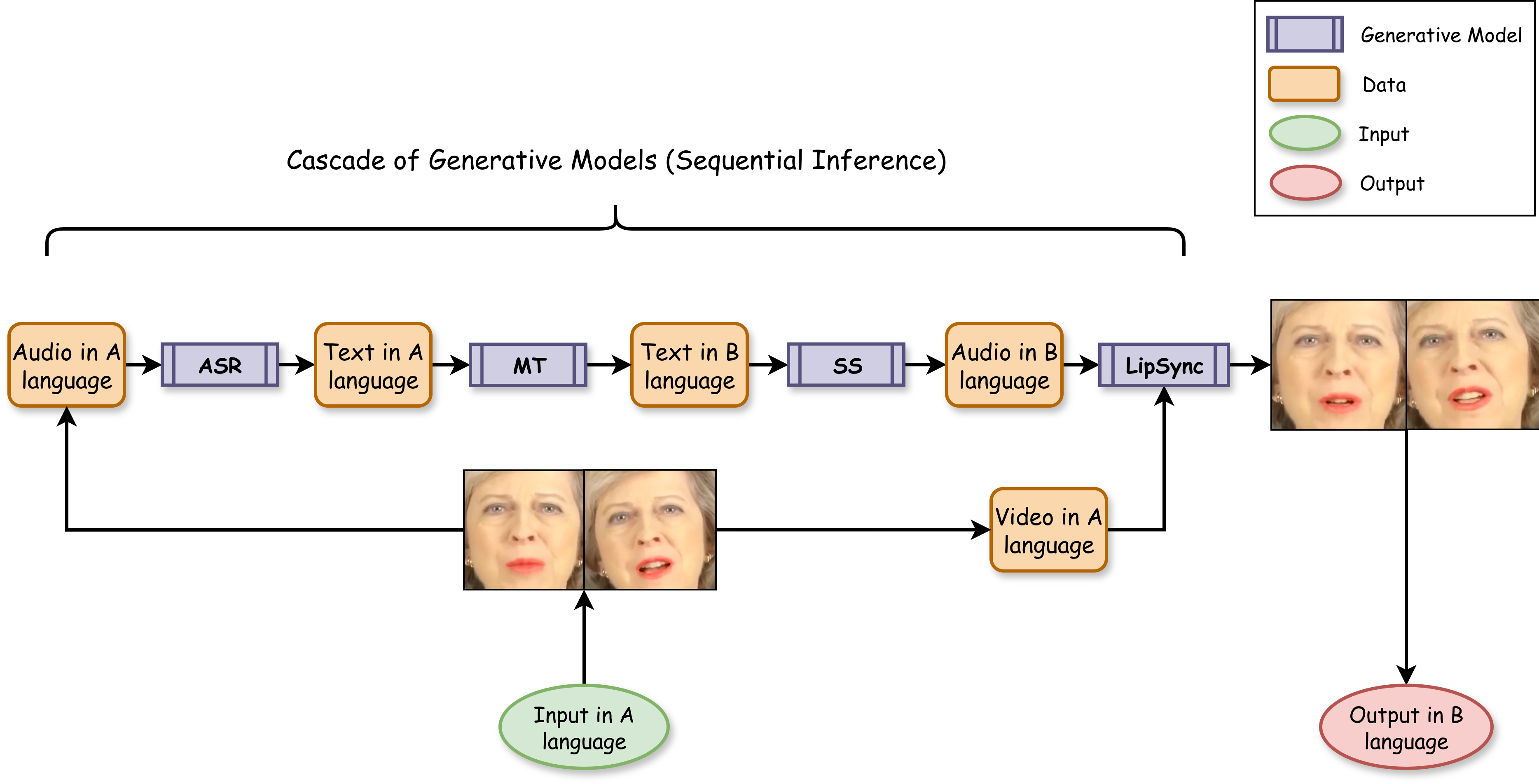}
    \caption{The sequential four-stage pipeline for video translation. An input in Language A is transformed by models for Automatic Speech Recognition (ASR), Machine Translation (MT), Speech Synthesis (SS), and Lip Synchronization (LipSync) to generate a fully translated and visually synchronized output in Language B. }
    \label{fig:intro_pipeline}
\end{figure}

The convergence of powerful Generative Artificial Intelligence (GenAI) and the global ubiquity of digital communication platforms is fundamentally reshaping human interaction. GenAI models can create novel, high-fidelity content—including text, code \cite{doropoulos2025beyond,rafiei2024beyond,acosta2025automated,liu2024empirical}, audio, and video \cite{kim2024dancecaps,yan2025video,oskooei2025facial}—offering the potential to make online environments more immersive and functional. This technological shift, occurring alongside the widespread adoption of platforms like video conferencing systems, Augmented/Virtual Reality (AR/VR) \cite{partarakis2025training,todino2025bridging,johnson2024collabvr,jover2025creating,mohamed2025artificial}, and social networks \cite{jun2025patent,huitema2024investigating,wang2025research} presents a transformative opportunity to dismantle longstanding barriers to global communication, most notably those of language \cite{soares2025digital,leung2024large}.
Within this domain, ``Video Translation''—also known as Video-to-Video or Face-to-Face Translation—represents an emerging paradigm of significant interest \cite{rafiei2025whisper}. Video translation aims to deliver a seamless multilingual experience by holistically translating all facets of human expression. This process involves converting spoken words, preserving the speaker's vocal tone and style, and critically, synchronizing their lip movements with the translated speech. Such a comprehensive translation fosters more natural and fluid conversations, providing immense value to international business, global academic conferences, and multicultural social engagements. Achieving this requires end-to-end pipelines that integrate multiple GenAI models for tasks such as automatic speech recognition (ASR), machine translation (MT), text-to-speech (TTS) synthesis, and lip synchronization (LipSync), as illustrated in Figure~\ref{fig:intro_pipeline}.

% ============ CASCADE DIAGRAM =============
% ========================================================

However, the practical deployment of these complex, multi-stage pipelines in real-time, large-scale applications is hampered by formidable system-level engineering challenges that have not been adequately addressed in existing research. GenAI models are computationally intensive, necessitating high-performance hardware like GPUs for timely execution. This requirement is magnified in real-time environments like video conferencing, giving rise to two primary bottlenecks:
\begin{enumerate}
%MDPI: Please confirm if the bold formatting is necessary; if not, please remove it. The following highlights are the same.
%AUTHOR: Removed.
\item Latency: 
The sequential execution of multiple deep learning models introduces significant processing delays. Each stage in the cascade adds to the total inference time, making it difficult to achieve the low-latency throughput required for smooth, uninterrupted conversation.
\item Scalability: In a multi-user video conference, a naive implementation would require each participant to concurrently process video streams from all other speakers. This approach results in a computational complexity of O(N²), which is prohibitively expensive and technically unmanageable, even for a small number of participants~(N).
\end{enumerate}

%MDPI: Please confirm if this need indent.  
%AUTHOR: Confirmed.
While many studies focus on improving individual models (model-level optimization), a distinct research gap exists in developing architectural and protocol-level solutions that enable these systems to perform effectively in real-time, multi-user settings (system-level optimization). {While architectural primitives like turn-taking and batching are known in distributed systems, their specific synthesis and application to solve the unique computational and semantic challenges of real-time generative video translation remains an open and critical research area.}
To address this critical gap and facilitate the practical deployment of video translation and similar GenAI pipelines, this paper introduces a system-level framework. This work makes the following primary contributions:
\begin{itemize}
\item We introduce a new \textit{System Architecture} designed to enable the scalable deployment of generative pipelines in multi-user, video-conferencing environments. The key innovations of this architecture are:
    \begin{itemize}
    \item A \textit{Token Ring }mechanism for managing speaker turns, which reduces the system's computational complexity from O(N²) to a linear O(N), thereby ensuring scalability.
    \item A \textit{Segmented Batched Processing} protocol with inverse throughput thresholding, which provides a mathematical framework for managing latency and achieving near real-time performance.
    \end{itemize}
\item We provide an empirical validation of our framework through a \textit{Proof-of-Concept }implementation. This study includes a performance analysis across both commodity and enterprise hardware, and a subjective evaluation with user study to confirm the system's practical viability and user perception.
\end{itemize}

{This manuscript is a substantially extended version of our preliminary work presented at the 25th International Conference on Computational Science and Its Applications \cite{rafiei2025whisper}. While the conference paper introduced the proof-of-concept, this journal article provides a significant and novel contribution by establishing the complete theoretical and empirical foundation for the proposed framework. The key advancements in this work include: (1)~a formal, mathematical methodology with algorithmic pseudocode for our architecture and protocols; (2) a comprehensive, multi-tiered objective performance evaluation across commodity, cloud, and enterprise-grade hardware that empirically validates our real-time processing claims; and (3) a statistically robust subjective evaluation based on a tripled participant pool (N~=~30) with new metrics that confirm the user-acceptability of our system's core design.}

The rest of this paper is organized as follows. Section~\ref{sec:literature} reviews the literature on key technologies and real-time multimedia systems. Section~\ref{sec:methodology} explains our proposed system architecture. Section~\ref{sec:poc_setup} describes the proof-of-concept implementation and the experimental setup. Section~\ref{sec:results} presents and analyzes the detailed results from our technical and user-based evaluations. Section~\ref{sec:discussion} discusses the implications of our findings, and Section~\ref{sec:conclusion} concludes the paper.

%%%%%%%%%%%%%%%%%%%%%%%%%%%%%%%%%%%%%%%%%%%%%%%%%%%%%%%%%%%%%%%%%%%%%%%%%%%%%%%%%%%%
%%%%%%%%%%%%%%%%%%%%%%%%%%%%%%%%%%%%%%%%%%%%%%%%%%%%%%%%%%%%%%%%%%%%%%%%%%%%%%%%%%%%
%%%%%%%%%%%%%%%%%%%%%%%%%%%%%%%%%%%%%%%%%%%%%%%%%%%%%%%%%%%%%%%%%%%%%%%%%%%%%%%%%%%%
\section{Literature Review}\label{sec:literature}

The ability to seamlessly translate spoken language in real-time is a long-standing goal in human-computer interaction, and modern Speech-to-Speech Translation (S2S) systems represent a significant step towards achieving it. The conventional S2S framework operates as a cascaded pipeline, beginning with an Automatic Speech Recognition (ASR) module to transcribe speech into text. This text is then translated into a target language by a Machine Translation (MT) component, and finally vocalized by a Speech Synthesis (SS) or Text-to-Speech (TTS) system. Each stage of this pipeline has been profoundly enhanced by deep learning. Neural ASR architectures have achieved robust performance even in challenging acoustic conditions \cite{malik2021automatic,prabhavalkar2023end,kheddar2024automatic}, while Neural Machine Translation (NMT) has become the standard, far exceeding the quality of earlier statistical methods \cite{nllb2024scaling,wang2022progress}. The emergence of Large Language Models (LLMs) has further advanced the state-of-the-art, enabling more contextually nuanced translations and even end-to-end multilingual capabilities \cite{rafiei2024beyond,zhu2023multilingual,oskooei2025repository}. In parallel, neural TTS systems can now generate highly natural and expressive speech, with zero-shot voice cloning techniques allowing for the preservation of a speaker's unique vocal identity \cite{casanova2024xtts,li2022unet}. These concurrent advances have established a strong foundation for high-fidelity audio translation, with multimodal generative AI now being explored for complex conversational simulations in fields like medical education \cite{chu2024synthetic}.

For video-based communication, audio translation alone is insufficient. Achieving a truly immersive experience requires synchronizing the speaker's lip movements with the translated audio, a task handled by Talking Head Generation or Lip Synchronization models. A critical requirement for these models within a video translation pipeline is language independence, ensuring that the visual output is driven purely by the audio phonemes rather than the semantics of a specific language. Foundational work in this area demonstrated the feasibility of language-agnostic lip-sync using neural networks, setting the stage for subsequent research \cite{kr2019towards}. Early successes were dominated by Generative Adversarial Networks (GANs), which proved effective at generating realistic facial textures and movements \cite{prajwal2020lip,bao2024milg,kim2021lip,vougioukas2019end,das2020speech,yin2022styleheat,hong2022depth}. More recently, the field has seen a paradigm shift towards Diffusion-based models, which often yield higher-quality and more stable \mbox{results \cite{lin2025omnihuman,li2024latentsync,xu2024vasa,shen2023difftalk,stypulkowski2024diffused}}, and models based on Neural Radiance Fields (NeRFs), which excel at creating photorealistic, view-consistent 3D talking heads \cite{ye2023geneface++,sun2023vividtalk,ye2024mimictalk,ye2024real3d,shin2024wav2nerf,song2025multi}. Beyond mere synchronization, research has also explored enhancing the expressiveness of these models, for instance, by enabling control over the emotional affect of the generated face \cite{zhao2022emotionally}. The performance of these systems is evaluated using a wide range of metrics, from visual fidelity to the analysis of underlying multimodal signals, such as the neural correlates of lip-sync imagery \cite{naebi2023performance}. The generalizability of these models across diverse languages remains a key area of investigation, confirming its importance for global applications \cite{rafiei2024can}.

While individual component models are well-researched, the integration of these parts into complete, end-to-end Video Translation systems is a less explored domain. Early, pre-neural attempts demonstrated the concept \cite{ritter1999face}, while later work discussed the potential of face-to-face translation without providing a full system implementation for real-world use~\cite{kr2019towards}. Some studies have integrated multilingual TTS with lip-syncing but did not focus on the practical challenges of deployment \cite{song2022talking}. A notable recent approach, TransFace, proposed an end-to-end model to avoid cascading separate modules \cite{cheng2023transface}. However, such end-to-end systems can sacrifice the modularity needed for independent component upgrades and fine-grained control. {Our own prior work has focused on achieving low-latency performance but did not address the broader system-level scalability challenges inherent in multi-user applications.}

{A critical synthesis of this literature reveals a distinct trajectory: while the individual component technologies for video translation have achieved remarkable maturity, the focus on holistic, deployable systems
%MDPI: Please confirm if the italics are necessary; if not, please remove them. The following highlights are the same.
%AUTHOR: Removed
 remains underdeveloped. Research has produced highly effective models for each stage of the pipeline, from robust ASR and massively multilingual MT to high-fidelity, zero-shot TTS \cite{kheddar2024automatic, nllb2024scaling, casanova2024xtts}. Similarly, the field of lip synchronization has rapidly evolved from foundational GAN-based methods to more photorealistic Diffusion and NeRF-based models \cite{prajwal2020lip, shen2023difftalk, ye2023geneface++}. However, integrating these components into a functional whole presents significant engineering challenges that are often addressed in isolation. For instance, some end-to-end systems sacrifice the modularity required to easily upgrade these rapidly evolving components \cite{cheng2023transface}, while other system-level work has focused primarily on mitigating latency for a single user without addressing the critical challenge of multi-user scalability \cite{rafiei2024seeing}. This creates a clear research gap: the need for a comprehensive, system-level framework that is both modular and explicitly designed to solve the dual, interconnected problems of real-time latency and quadratic scalability inherent in any practical, multi-user deployment.}

Despite these advances in generative models, a significant research gap persists in the system-level engineering required for deploying these pipelines in real-time, multi-user environments like video conferencing. This paper directly addresses this gap by proposing and empirically validating a novel framework designed specifically to make real-time video translation feasible, scalable, and efficient for real-world deployment.

%%%%%%%%%%%%%%%%%%%%%%%%%%%%%%%%%%%%%%%%%%%%%%%%%%%%%%%%%%%%%%%%%%%%%%%%%%%%%%%%%%%%
%%%%%%%%%%%%%%%%%%%%%%%%%%%%%%%%%%%%%%%%%%%%%%%%%%%%%%%%%%%%%%%%%%%%%%%%%%%%%%%%%%%%
%%%%%%%%%%%%%%%%%%%%%%%%%%%%%%%%%%%%%%%%%%%%%%%%%%%%%%%%%%%%%%%%%%%%%%%%%%%%%%%%%%%%
\section{Methodology}\label{sec:methodology}

The primary contribution of this work is a comprehensive, system-level framework designed to bridge the gap between the potent capabilities of generative AI models and their practical application in real-time, multi-user communication systems. Our methodology is not focused on model-level optimization but rather on the architectural and protocol-level engineering required to make such systems feasible, scalable, and efficient. This section provides a detailed exposition of this framework, beginning with the foundational generative pipeline that serves as our testbed, followed by an in-depth analysis of our core architectural and protocol-level innovations.

\subsection{GenAI Pipeline}\label{subsec:pipeline}

To empirically validate our system-level framework, we first implemented a modular, end-to-end pipeline for the task of video translation. The choice of a modular design is deliberate; it ensures that each functional component can be independently upgraded as more advanced models become available, thereby future-proofing the architecture. The pipeline's objective is to process an input video of a speaker and generate a semantically equivalent output where the speech is translated and the lip movements are synchronized to the new audio, while preserving the original vocal identity. As illustrated in Figure~\ref{fig:intro_pipeline}, the pipeline comprises four discrete, sequential stages:

\begin{enumerate}
    \item Automatic Speech Recognition (ASR): %MDPI: Please confirm if the bold formatting is necessary; if not, please remove it. The following highlights are the same.
    %AUTHOR: Removed
    The pipeline ingests the raw audio stream from the input video. An ASR model performs transcription, converting the spoken phonemes into a textual representation in the source language.
    \item Machine Translation (MT): The source-language text is then passed to a multilingual MT model. This component is responsible for translating the text to the desired target language while preserving the original meaning and context.
    \item Speech Synthesis (SS): A Text-to-Speech (TTS) model, critically equipped with zero-shot voice cloning capabilities, synthesizes the translated text into an audio waveform. The voice cloning functionality is essential for preserving the speaker's unique vocal timbre and prosody, which is paramount for maintaining identity and providing a naturalistic user experience.
    \item Lip Synchronization (LipSync): Finally, a language-agnostic LipSync model receives two inputs: the original, silent video frames and the newly synthesized target-language audio. It then generates a new video by modifying the speaker's mouth region to synchronize precisely with the translated audio, completing the video translation process.
\end{enumerate}

\subsection{System-Level Challenges}\label{subsec:challenges}

A naive implementation of the above pipeline within a video conferencing application would fail due to two fundamental and prohibitive system-level challenges. A core part of our methodology is to first formally define these challenges to motivate our \mbox{subsequent solutions.}

\begin{itemize}
    \item Challenge 1: Cumulative Latency. The sequential, cascaded execution of four deep learning models results in a significant cumulative processing delay. The total inference time is the sum of the latencies of each stage ($p_{\text{total}} = p_{\text{ASR}} + p_{\text{MT}} + p_{\text{SS}} + p_{\text{LipSync}}$). In real-time communication, which is perceptually sensitive to delays exceeding a few hundred milliseconds, this cumulative latency leads to an unacceptable Quality of Service (QoS) degradation, destroying conversational fluidity.
    \item Challenge 2: Quadratic Scalability. In a multi-user environment with $N$ participants, a brute-force architecture would require each of the $N$ participants to maintain $N-1$ concurrent processing streams to translate video from all other speakers. This creates a demand for $N \times (N-1)$ total processing instances, resulting in a computational complexity of $\mathcal{O}(N^2)$. This quadratic growth is computationally intractable and economically unviable, making the system unscalable beyond a trivial number of~users.
\end{itemize}

\subsection{Proposed System Architecture}\label{subsec:architecture}

\begin{figure}[H]
   % \centering
    \includegraphics[width=1.0\textwidth]{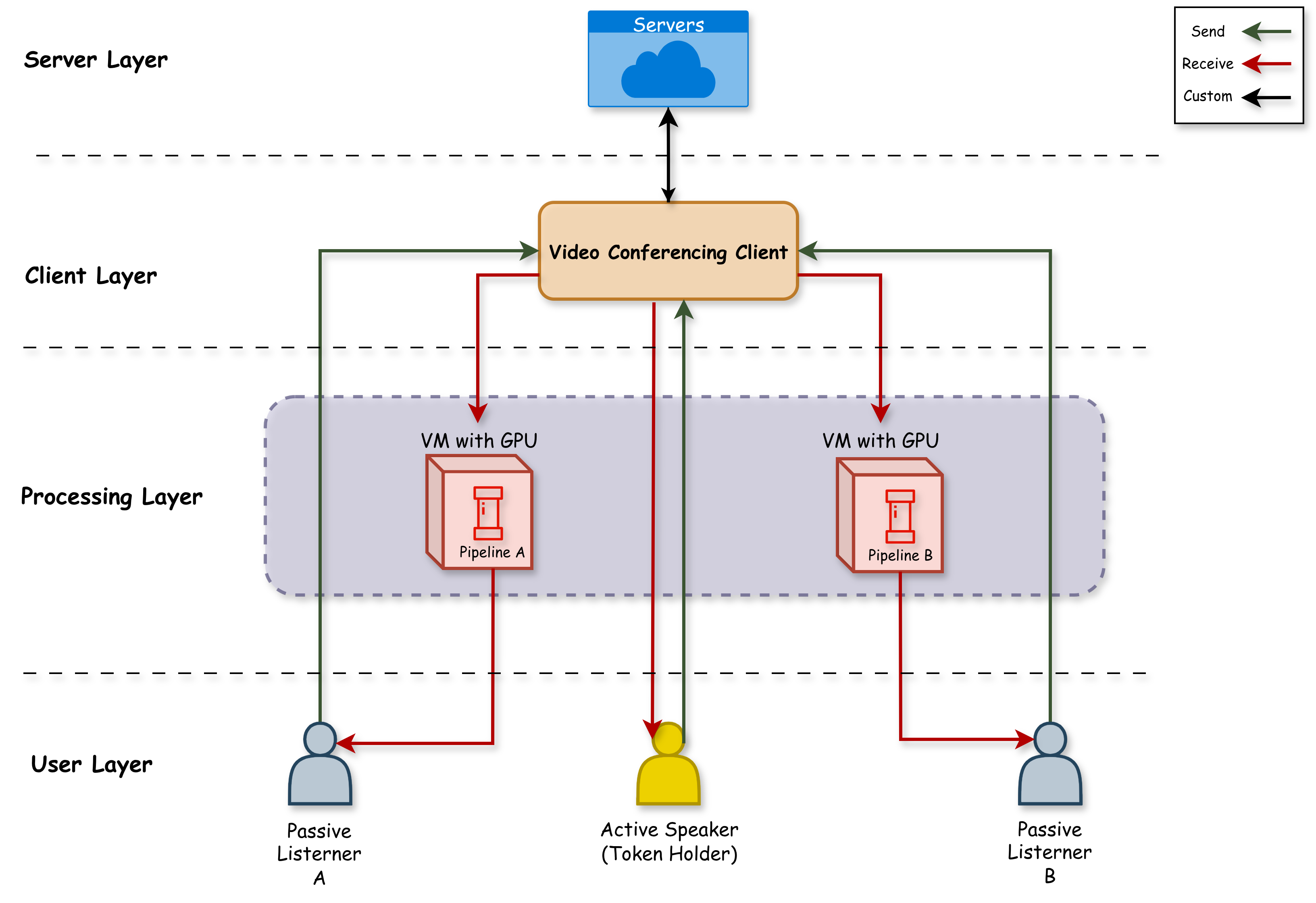}
    \caption{The proposed four-layer system architecture. This design decouples the user-facing client from the back-end Processing Layer, which acts as an orchestration engine for a pool of GPU resources.}
    \label{fig:system_architecture}
\end{figure}

Our solution begins with a robust system architecture based on the principle of \textit{decoupling}. We separate the user-facing application logic from the intensive computational workload. This design pattern enhances modularity, maintainability, and scalability. As shown in Figure~\ref{fig:system_architecture}, the architecture comprises four distinct layers:

\begin{itemize}
    \item Server Layer: This layer represents the core video conferencing infrastructure (e.g., signaling servers) responsible for session management and establishing communication channels between clients.
    \item Client Layer: This is the user-facing application (e.g., any WebRTC-based video conferencing client). Its responsibilities are to manage user interaction, handle media stream encoding/decoding, and render the final video.
    \item Processing Layer: This intermediary orchestration engine is the core of our solution. It is a stateless service that manages a dynamic pool of GPU resources. It intercepts media streams, allocates pipeline instances on demand, and routes the translated output to the correct recipients.
    \item User Layer: This abstract layer represents the human participants who interact with the Client Layer and specify their desired target language for translation.
\end{itemize}

% ========== FIGURE PLACEHOLDER ==========
% ======================================

\subsection{The ``Token Ring'' Mechanism}\label{subsec:token_ring}

To solve the quadratic scalability problem defined in Section~\ref{subsec:challenges}, we introduce a novel turn-taking protocol we term the ``Token Ring'' mechanism. This protocol is not merely a heuristic but a structured approach to resource management that fundamentally alters the system's computational complexity.

\subsubsection{Formal Cost Modeling and Complexity Analysis}
To provide a rigorous justification for this mechanism, we first define our terms and then formally model the computational cost of the system.

\paragraph{Definitions}%MDPI: Please confirm whether we should retain it as a level 4 heading (the colon needs to be removed), or suggest changing it to paragraph format (the colon can be retained). The following highlights are the same.
%AUTHOR: We should retain it as a level 4 heading 
\begin{itemize}
    \item Let $N \in \mathbb{Z}^+$ be the total number of participants in the meeting, where $N \ge 2$.
    \item Let $C$ be the constant representing the computational cost (e.g., GPU resource allocation) of a single, complete video translation pipeline instance.
    \item Let $P$ be the total system cost, defined as the aggregate cost of all concurrently running pipeline instances.
\end{itemize}

\paragraph{Scenario 1: The Naive (Brute-Force) System}%MDPI: Please confirm whether we should retain it as a level 4 heading (the period needs to be removed), or suggest changing it to paragraph format (the period can be retained). The following highlights are the same.
%AUTHOR: We should retain it as a level 4 heading 
In an unmanaged system, each of the $N$ participants must have the capacity to process incoming streams from all other $N-1$ participants. The total system cost, $P_{\text{naive}}$, is therefore the product of the number of participants and the number of streams each must process:
\begin{equation}
    P_{\text{naive}} = C \cdot N \cdot (N-1) = C(N^2 - N)
\end{equation}%MDPI: Please confirm if this need indent. The following highlights are the same.
%AUTHOR: Needs indent.

As $N \to \infty$, the cost is dominated by the quadratic term, establishing a computational complexity of $\mathcal{O}(N^2)$. This quadratic growth renders the system computationally intractable and economically unviable for any non-trivial number of participants.

\paragraph{Scenario 2: The Generalized Token Ring System}
Our proposed mechanism designates one participant as the \textit{Active Speaker} and the remaining $N-1$ as \textit{Passive Listeners}. The key insight is that the system cost is no longer a function of the total number of participants, but of the diversity of target languages~requested.

\paragraph{Generalized Model}
\begin{itemize}
    \item Let $L$ be the set of unique target languages selected by the $N-1$ passive listeners.
    \item Let $k = |L|$ be the cardinality of this set, representing the number of distinct target languages.
    \item The value of $k$ is bounded such that $1 \le k \le N-1$.
\end{itemize}
The total number of required pipeline instances is now equal to $k$, as all listeners requesting the same target language can be served by a single, shared pipeline instance. The total system cost, $P_{\text{token}}$, is therefore:
\begin{equation}
    P_{\text{token}} = C \cdot k, \quad \text{where } k \in [1, N-1]
\end{equation}

\paragraph{Complexity Analysis of Boundary Conditions}
\begin{itemize}
    \item Worst-Case Complexity: The worst case occurs when every passive listener selects a unique target language. In this scenario, $k = N-1$. The cost becomes $P_{\text{token}}^{\text{worst}} = C \cdot (N-1)$, establishing a clear upper bound with a linear computational complexity of $\mathcal{O}(N)$.
    \item Best-Case Complexity: The best case occurs when all passive listeners select the same target language. In this scenario, $k = 1$. The cost becomes $P_{\text{token}}^{\text{best}} = C$, which is a constant cost independent of the number of users. This establishes a lower bound with a complexity of $\mathcal{O}(1)$. Using Asymptotic notation, we can state the best-case complexity is also $\Omega(1)$, indicating a constant time complexity.
\end{itemize}
This formal analysis demonstrates that the Token Ring mechanism transforms an intractable $\mathcal{O}(N^2)$ problem into a highly manageable linear problem, which in many practical scenarios trends towards a constant-time $\mathcal{O}(1)$ solution.

{To formalize the operational logic of the Token Ring mechanism, we present its core orchestration algorithm in Algorithm~\ref{alg:token_ring}. This algorithm executes within the Processing Layer upon any change in the active speaker, managing the allocation, reuse, and deallocation of GPU pipeline instances to maintain a minimal computational footprint.}

\begin{algorithm}[H]
\caption{Token Ring Stream Orchestration Protocol}
\label{alg:token_ring}
\begin{lstlisting}
Input: P = {p1, ..., pN} participants, s speaker, G GPU pool
State: Pi: L -> G  (language to pipeline mapping)
Output: Updated routing and allocation

function UpdateOrchestration(P, s, G):
  Route(s.video, s.id, bypass=true)   # speaker bypasses processing

  L <- {p.lang | p in P - {s}}        # required languages
  k <- |L|                            # count: k in [0, N-1]

  for each ell in L:                  # allocate new pipelines
    if ell not in dom(Pi):
      if G not empty:
        g <- Alloc(G)
        Init(g, ell, s.lang)
        Pi(ell) <- g
      else:
        LogError(ell); continue

  for each p in P - {s}:              # route to listeners
    if p.lang in dom(Pi):
      g <- Pi(p.lang)
      Route(s.video, g.in); Route(g.out, p.id)

  Ls <- dom(Pi) - L                   # deallocate stale pipelines
  for each ell in Ls:
    g <- Pi(ell)
    Decommission(g); Dealloc(g, G)
    Pi <- Pi - {(ell, g)}

  assert P = C * |dom(Pi)| = C * k    # O(N) cost
\end{lstlisting}
\end{algorithm}

Algorithm~\ref{alg:token_ring} implements the Token Ring mechanism, executing on each speaker transition. Three invariants ensure correctness: (1) speaker $s$ receives bypass stream (no self-translation), (2) $L$ determines minimal pipeline allocation (no redundancy), and (3) stale deallocation (lines 20-24) maintains $|\text{dom}(\Pi)| = k \leq N-1$, preserving $\mathcal{O}(N)$ complexity from Section~\ref{subsec:token_ring}. The key insight: transform from participant-centric allocation ($N \times (N-1)$ instances) to language-centric ($k$ instances only). With linguistic homogeneity (e.g., bilingual meetings), this approaches $\mathcal{O}(1)$ as $k \ll N$.

\subsubsection{Justification and Design Rationale}%MDPI: We added the section number. Please confirm.
%AUTHOR: Confirmed
The enforcement of a single active speaker may initially appear to be a constraint on the natural, fluid dynamics of conversation where interruptions and overlapping speech can occur. However, this design choice is a deliberate and necessary trade-off made to ensure intelligibility in a translated context. In the target domain of professional, educational, or formal multilingual communication, effective information exchange already relies on clear, sequential turn-taking. Simultaneous speech from multiple parties, when translated into multiple languages, would result in a cacophony of audio streams, rendering the conversation unintelligible for all listeners. Therefore, the Token Ring mechanism does not impose an unnatural constraint; rather, it formalizes an existing social protocol for coherent communication and leverages it as an architectural cornerstone for achieving computational tractability and ensuring a high Quality of Experience (QoE).

\subsection{Segmented Batched Processing}\label{subsec:segmented_processing}

To solve the latency problem, we developed a protocol that manages the high intrinsic processing delay of the pipeline to deliver a perceptually real-time user experience. Our analysis deliberately isolates computational latency from network latency, a standard practice for modeling system performance.

\subsubsection{System Performance Characterization}
Our protocol is based on a rigorous characterization of the pipeline's performance.

\paragraph{Definitions}%MDPI: Please confirm whether we should retain it as a level 4 heading (the colon needs to be removed), or suggest changing it to paragraph format (the colon can be retained). 
%AUTHOR: We should retain it as a level 4 heading 
\begin{itemize}
    \item Let $t \in \mathbb{R}^+$ be the duration of an input video segment in seconds.
    \item Let $p: \mathbb{R}^+ \to \mathbb{R}^+$ be the function describing the pipeline's total processing time for a segment of duration $t$.
\end{itemize}

Through empirical analysis, we model the behavior of $p(t)$ as a piecewise function defined by a hardware-dependent constant, the \textit{System Threshold (}$T_{thresh}$). This threshold marks the transition point between two operational regimes:
\begin{equation}
p(t) = 
\begin{cases} 
      p(t) > t & : t < T_{thresh} \quad \text{(System Lag Regime)} \\
      p(t) \le t & : t \ge T_{thresh} \quad \text{(Real-Time Regime)}
   \end{cases}
\label{eq:segmented}
\end{equation}
In the System Lag Regime, system overheads dominate, and the pipeline cannot keep pace with real-time. In the Real-Time Regime, initial overheads are amortized, and the processing time $p(t)$ grows sub-linearly, a behavior we empirically model as $p(t) \approx \log(t)$.

\subsubsection{The Real-Time Viability Condition}
To operationalize this model, we define the \textit{Reciprocal Throughpu}t, $\tau(t)$, a dimensionless metric that normalizes processing time against real-time duration:
\begin{equation}
    \tau(t) = \frac{p(t)}{t}
    \label{eq:reciprocal}
\end{equation}
The value of $\tau(t)$ directly indicates system viability: $\tau(t) > 1$ implies the system is falling behind, while $\tau(t) \le 1$ implies it is at or ahead of real-time. Continuous, uninterrupted playback is only possible if the system operates consistently in the $\tau(t) < 1$ state. This leads us to our core operational requirement:
\begin{equation}
    \textbf{Condition for Smooth Playback:} \exists T \text{ such that } \tau(t) < 1.0 \text{ for all } t \ge T
    \label{eq:smooth}
\end{equation}
We define the \textit{Optimal Segment Duration ($T_{opt}$)} as the minimum value of $t$ that satisfies this condition, ensuring both stability and minimal initial latency.
\begin{equation}
    T_{opt} = \min(\{t \mid \tau(t) < 1.0\})
\end{equation}

\subsubsection{Overlapping Buffering}
Once $T_{opt}$ is determined for a given hardware configuration, the protocol operationalizes Equation~(\ref{eq:smooth}) by segmenting the input stream into fixed-duration chunks of length $T_{opt}$ and employing a strategy of overlapping buffering. This guarantees a continuous stream for the listener after a single, initial buffering event, transforming a system with high intrinsic latency into one that delivers a perceptually near real-time experience.

{The operational logic of the segmented processing protocol is detailed in Algorithm~\ref{alg:segmentation}. This algorithm runs on a dedicated GPU instance within the Processing Layer. It continuously reads segments of optimal duration ($T_{opt}$) from the active speaker's stream and processes them asynchronously, using a queue to buffer the output and ensure smooth, uninterrupted playback for the listener after an initial startup delay.}

\begin{algorithm}[H]
\caption{Segmented Processing with Overlapping Buffering Protocol}
\label{alg:segmentation}
\begin{lstlisting}
Input: S input stream, Phi pipeline, T segment duration
Output: S' output stream
Precondition: exists T such that tau(T) = p(T)/T < 1.0  (real-time viable)

function ProcessStream(S, Phi, T):
  Q <- empty queue          # FIFO job queue
  buffered <- false         # buffer state
  k <- 0                    # segment counter

  while S.HasData():
    vk <- Read(S, T)        # get segment of length T
    k <- k + 1
    thetak <- AsyncJob(Phi, vk)   # non-blocking call
    Enqueue(Q, thetak)

    if not buffered:        # initial buffering
      Wait(Front(Q))        # block on first job
      buffered <- true

    while Q not empty and Done(Front(Q)):   # stream completed jobs
      theta <- Dequeue(Q)
      Write(S', Result(theta))

  while Q not empty:        # drain remaining jobs
    theta <- Dequeue(Q)
    Wait(theta)
    Write(S', Result(theta))

  Close(S')
  assert tau(T) < 1.0 implies no buffering after startup
\end{lstlisting}
\end{algorithm}

Algorithm~\ref{alg:segmentation} operationalizes the Segmented Batched Processing protocol by implementing a producer-consumer pattern with explicit initial buffering. The algorithm's correctness depends critically on the precondition that $T_{\text{opt}}$ satisfies the Real-Time Viability Condition ($\tau(T_{\text{opt}}) < 1.0$), as established empirically in Section~\ref{sec:results}. The one-time blocking wait in lines 12-15 constitutes the ``predictable initial delay'' identified in our methodology, after which the overlapping mechanism ensures continuous output: while segment $v_i$ streams to the listener, segment $v_{i+1}$ is concurrently read from the input, and segment $v_{i+2}$ undergoes pipeline processing. This three-stage overlap is possible precisely because $p(T_{\text{opt}}) < T_{\text{opt}}$, meaning processing completes faster than real-time playback. The queue drain phase (lines 23-28) ensures graceful stream termination without data loss. Importantly, the total perceived latency remains bounded at $t_{\text{startup}} = p(T_{\text{opt}})$ regardless of stream duration, as the steady-state condition $\tau(T_{\text{opt}}) < 1.0$ eliminates all subsequent buffering events, thereby delivering the perceptually smooth experience validated in our subjective evaluation (\ref{subsec:subjective}).
%MDPI: The first citation of each table should appear in numerical order. Table 4 should be cited after Tables 1-3. Please confirm.
%AUTHOR: Confirmed

\subsubsection*{Justification and Design Rationale}
While the ideal for any real-time system is zero latency, the significant computational demands of cascaded generative models make this physically unattainable with current technology. The system designer is therefore faced with a critical engineering trade-off: (a)~attempt a continuous, low-latency stream that is highly susceptible to frequent stuttering, buffering, and desynchronization as the pipeline struggles to keep up, or (b) introduce a predictable, one-time latency at the start of a speaking turn in exchange for guaranteed smooth, uninterrupted playback thereafter. From a user experience (UX) and psycholinguistic perspective, the latter is vastly superior. Predictable, initial delays are quickly adapted to by users, whereas intermittent, unpredictable interruptions are highly disruptive and destroy the perception of conversational flow. Our protocol makes the deliberate choice to front-load this unavoidable computational cost, thereby ensuring a high-quality, reliable, and perceptually smooth experience for the duration of the speaker's turn.

% \begin{table}[H] 
% \caption{This table provides an overview of our dataset, collected for 3 culturally distinct languages}
% \begin{tabularx}{\textwidth}{{@{\hskip 1mm}C@{\hskip 1mm}C@{\hskip 1mm}C@{\hskip 1mm}C@{\hskip 1mm}}}
% \toprule
% \textbf{Language}	& \textbf{Origin}	& \textbf{Source}   & \textbf{Length (mins)} \\
% \midrule
% Turkish		& Turkic			& MediaSpeech    & 67   \\
% Arabic		& Semitic			& MediaSpeech    & 54   \\
% Persian		& Indo-Iranian		& Persian Speech Corpus    & 59   \\
% \bottomrule
% \end{tabularx}
% \label{tab:dataset}
% \end{table}

% \begin{figure}[H]
% \begin{adjustwidth}{-\extralength}{0cm}
% \centering
% \includegraphics[width=15.5cm]{images/Paper_workflow.png}
% \end{adjustwidth}
% \caption{A Modular Face-to-Face Translation Workflow.\label{fig:workflow}}
% \end{figure}  

% ===============================================================
% ===============================================================
% ===============================================================

\section{Proof-of-Concept \& Experimental Setup}\label{sec:poc_setup}

This section details the empirical framework designed to validate the theoretical methodology presented in Section~\ref{sec:methodology}. The primary objective is not to build a production-ready, commercial-grade application, but rather to develop a simplified yet functional \textit{Proof-of-Concept (PoC)}. This PoC serves two crucial purposes: first, as a testbed for rigorously characterizing the performance of our proposed architecture and protocols under controlled conditions; and second, to demonstrate the feasibility of our approach. By maintaining simplicity and leveraging widely available technologies, this PoC is intended to provide a foundational, universal framework that developers and businesses can build~upon.

\subsection{Software}
\label{subsec:software}

The PoC was implemented as a complete system, encompassing both the back-end generative pipeline and a front-end user interface to simulate a real-world video conferencing environment.

\subsubsection{Models \& Libraries}
The four-stage video translation pipeline described in Section~\ref{subsec:pipeline} was implemented using state-of-the-art, open-source models selected for their performance and accessibility. The modular design allows for each component to be independently benchmarked and updated. To ensure full reproducibility---a cornerstone of high-quality scientific \mbox{research---the} exact models and libraries, along with their version numbers, are detailed in Table~\ref{tab:reproducibility}.

\begin{table}[H]
  %  \centering
    \caption{Model and Library Specifications for Reproducibility.}
    \label{tab:reproducibility}
    \setlength{\tabcolsep}{9.7mm}
    \begin{tabular}{lll}
        \toprule
        \textbf{Stage} & \textbf{Model/Library} & \textbf{Version} \\
        \midrule
        ASR & OpenAI Whisper & 3.1.1 \\
        MT & Meta SeamlessM4T & 1.0.0 \\
        SS (TTS) & Coqui XTTS & 2.0.2 \\
        LipSync & Wav2Lip (GAN) & 1.0.0 \\
        \midrule
        \textbf{Framework} & \textbf{Library} & \textbf{Version} \\
        \midrule
        Deep Learning & PyTorch & 2.1.0 \\
        GPU Acceleration & CUDA Toolkit & 12.1 \\
        Programming Language & Python & 3.10 \\
        \bottomrule
    \end{tabular}
\end{table}

\subsubsection{Graphical User Interface}
To facilitate subjective evaluation and simulate a realistic user experience, we developed a simple web-based user interface (UI) using React.js, CSS, and HTML. The UI allows users to initiate a simulated video call, select a target language for translation from a dropdown menu, and view the final translated video with synchronized lip movements. Real-time signaling and communication between clients were managed using Socket.io, while peer-to-peer video and audio streaming were implemented with the Simple-Peer (WebRTC) library.

Crucially, the prototype was designed to be platform-agnostic. This ensures that the core processing logic can be integrated into any existing video conferencing platform with minimal modifications, reinforcing the universality of our proposed architecture. The simplicity of the UI is deliberate, focusing squarely on testing the feasibility and perceptual quality of the back-end system.

\subsection{Hardware}
\label{subsec:hardware}

A key objective of this study is to characterize our system's performance across a spectrum of computational capabilities, from consumer-grade hardware to enterprise-level infrastructure. This approach demonstrates the accessibility of our framework while also validating its performance in a production-like environment. To this end, we established three distinct hardware testbeds:

\begin{itemize}
    \item Commodity-Tier (Commercial Laptop): An NVIDIA RTX 4060 Laptop GPU was used to represent a typical high-performance consumer or developer machine. Experiments on this tier were run on a local machine.
    \item Cloud/Datacenter-Tier (Google Colab): An NVIDIA T4 GPU, a widely available and cost-effective datacenter card, was used to simulate a common cloud computing environment. The T4 represents a conservative baseline for cloud performance.
    \item Enterprise-Tier (Google Colab Pro): An NVIDIA A100 GPU, a high-performance card designed for demanding AI workloads, was used to represent an enterprise-grade production environment.
\end{itemize}

This multi-tiered hardware strategy allows us to directly test the hardware-dependent nature of the system threshold ($T_{thresh}$) and validate the real-time viability of our protocol as defined in Section~\ref{subsec:segmented_processing}.

\subsection{Dataset}
\label{subsec:dataset}

For the objective evaluation, we constructed a standardized test dataset to ensure consistent and comparable measurements. The dataset consists of 8-s video clips of speakers from diverse linguistic backgrounds, sourced from public-domain interviews. To systematically analyze the relationship between input duration and processing time ($p(t)$), these 8-s samples were meticulously segmented into clips of five distinct durations: 1, 2, 3, 5, and 8 s. (For all objective inference time analyses, a single benchmark scenario---translating a German-language video into English---was used to provide a consistent performance profile and eliminate language-pair variability as a confounding~factor.) %MDPI: Footnotes are not supported in our journal. We have therefore included this paragraph in the main text. Please confirm.
%AUTHOR: Confirmed

\subsection{Evaluation}
\label{subsec:eval_protocol}

Our evaluation protocol is divided into two complementary components: an objective analysis of computational performance and a subjective assessment of user experience. All details of the evaluation methodology are consolidated within this section to provide a clear and comprehensive overview of our experimental procedures.

\subsubsection{Objective Evaluation}%MDPI: We revised this to heading level 3 and added the number. Please confirm. The following highlight is the same.
%AUTHOR: Confirmed
The goal of this evaluation is to empirically measure the performance metrics defined in our methodology (Section~\ref{subsec:segmented_processing}). We measured two primary metrics: (1) \textit{Inference Time} ($p(t)$), the wall-clock time in seconds required for the entire pipeline to process a video segment of duration $t$, and (2) \textit{Reciprocal Throughput} ($\tau(t)$), calculated as per Equation~(\ref{eq:reciprocal}) ($\tau(t) = p(t)/t$). The inference time was recorded for the full pipeline and for each individual module to identify bottlenecks. To ensure statistical robustness and account for minor variations, each experiment on each hardware testbed and for each segment duration was iterated three times. The final reported values are the mean and standard deviation of these three runs.

\subsubsection{Subjective Evaluation}
The goal of this evaluation is to quantify the perceptual quality of the system's output from an end-user perspective. A pool of 30 participants was recruited to ensure the statistical robustness of our findings. {The study involved 30 volunteer participants (18~male, 12 female) aged between 22 and 48 (mean age: 31.5). Participants were recruited from diverse geographical and linguistic backgrounds, including North America, Europe, and the Middle East, ensuring a range of native speakers for languages including English, German, and Turkish. All participants reported high familiarity with standard video conferencing tools but possessed varying levels of expertise in generative AI, representing a general user population rather than a panel of AI experts. This demographic information is provided to improve the transparency and generalizability of our findings.}

Participants interacted with the PoC's user interface, where they were shown translated video clips generated under the specific hardware conditions detailed in Section~\ref{sec:results}. They were then asked to rate their experience based on five criteria using a 5-point Likert scale (1~=~Bad, 2~=~Poor, 3~=~Fair, 4~=~Good, 5~=~Excellent). The aggregated results were analyzed using the Mean Opinion Score (MOS) protocol. The five evaluation criteria were designed to provide a holistic assessment of the user experience, as detailed in Table~\ref{tab:subjective_criteria}.

\begin{table}[H]
  %  \centering
    \caption{Criteria for Subjective User Experience Evaluation.}
    \label{tab:subjective_criteria}
    \begin{tabularx}{\textwidth}{llX}
        \toprule
        \textbf{Metric} & \textbf{Abbr.} & \textbf{Description} \\
        \midrule
        Lip Sync Accuracy & LSA & How accurately do the speaker's lip movements match the translated audio? \\
        Motion Naturalness & MN & How natural and realistic are the generated lip and mouth movements, regardless of sync? \\
        Visual Quality & VIQ & What is the overall visual quality of the video, including any artifacts or blurring in the mouth region? \\
        Vocal Quality & VOQ & How natural is the synthesized voice, and how well does it preserve the original speaker's tone? \\
        Startup Delay Acceptability & SDA & Considering the benefit of a smooth, uninterrupted video, how acceptable was the initial delay before playback began? \\
        \bottomrule
    \end{tabularx}
\end{table}

This comprehensive experimental setup, encompassing a functional PoC, multi-tiered hardware, and a rigorous dual-pronged evaluation protocol, provides the empirical foundation for the results and analysis presented in the following section.

% ===============================================================
% ===============================================================
% ===============================================================
  
\section{Results \& Analysis}\label{sec:results}

This section presents the empirical findings from the comprehensive evaluation protocol detailed in Section~\ref{sec:poc_setup}. The results are organized into two main parts. First, we present the objective performance analysis, which serves to empirically validate the theoretical models of our system architecture and processing protocols. Second, we present the subjective user experience analysis, which quantifies the perceptual quality of the proof-of-concept and confirms its practical viability from an end-user perspective.

\subsection{Objective Performance Analysis}

The objective evaluation was designed to rigorously characterize the computational performance of the video translation pipeline across the three distinct hardware tiers. The primary goals were to: (1) validate the piecewise performance model and sub-linear scaling behavior of the pipeline's inference time, $p(t)$; and (2) empirically determine the conditions under which the ``Real-Time Viability Condition'' ($\tau(t) < 1.0$) is met.

Table~\ref{tab:objective_results} presents the full, aggregated results for total pipeline inference time ($p(t)$) and the calculated Reciprocal Throughput ($\tau(t)$) for each hardware testbed across the five video segment durations.

\begin{table}[H]
    %\centering
    \caption{Objective Performance Metrics Across Hardware Tiers. Inference Time ($p(t)$) is in seconds. Reciprocal Throughput ($\tau(t)$) is a dimensionless ratio. Values are presented as Mean $\pm$ Standard Deviation from three runs.}
    \label{tab:objective_results}
 \small  \setlength{\tabcolsep}{2.2mm}
  \begin{tabular}{lccc}
        \toprule
        \textbf{Hardware} & \boldmath{\textbf{Video Length ($t$)}} & \boldmath{\textbf{Inference Time ($p(t)$)}} & \boldmath{\textbf{Reciprocal Throughput ($\tau(t)$)}} \\
        \midrule
        \textbf{NVIDIA T4} %MDPI: 1. Please confirm if the bold formatting is necessary; if not, please remove it. The following highlights are the same. 2. Please state the name of the manufacturer, city, and country from where the equipment was sourced. The following highlights are the same.
        %AUTHOR: As mentioned in the text, they are provided by a a Google Colab notebook service.
        & 1~s & 8.99 $\pm$ 0.11 & 8.99 \\
        (Cloud Baseline) & 2~s & 10.27 $\pm$ 0.15 & 5.14 \\
        & 3~s & 10.92 $\pm$ 0.09 & 3.64 \\
        & 5~s & 12.01 $\pm$ 0.13 & 2.40 \\
        & 8~s & 12.70 $\pm$ 0.10 & 1.59 \\
        \midrule
        \textbf{NVIDIA RTX 4060} & 1~s & 4.52 $\pm$ 0.08 & 4.52 \\
        (Commodity) & 2~s & 4.81 $\pm$ 0.09 & 2.41 \\
        & 3~s & 5.10 $\pm$ 0.08 & 1.70 \\
        & 5~s & 5.68 $\pm$ 0.10 & 1.14 \\
        & 8~s & 6.55 $\pm$ 0.07 & \textbf{0.82} \\ %MDPI: Please confirm if the bold formatting is necessary; if not, please remove it. The following highlights are the same.
        %AUTHOR: They are necessary
        \midrule
        \textbf{NVIDIA A100} & 1~s & 1.87 $\pm$ 0.04 & 1.87 \\
        (Enterprise) & 2~s & 2.08 $\pm$ 0.05 & 1.04 \\
        & 3~s & 2.29 $\pm$ 0.03 & \textbf{0.76} \\
        & 5~s & 2.71 $\pm$ 0.05 & \textbf{0.54} \\
        & 8~s & 3.34 $\pm$ 0.04 & \textbf{0.42} \\
        \bottomrule
    \end{tabular}
\end{table}

The raw data clearly demonstrates two trends. First, for any given video length, the inference time scales inversely with the computational power of the GPU, with the A100 being significantly faster than the RTX 4060, which is in turn faster than the T4. Second, for all hardware, the increase in inference time is sub-linear with respect to the increase in video length.

\begin{figure}[H]
  %  \centering
    \includegraphics[width=1.0\textwidth]{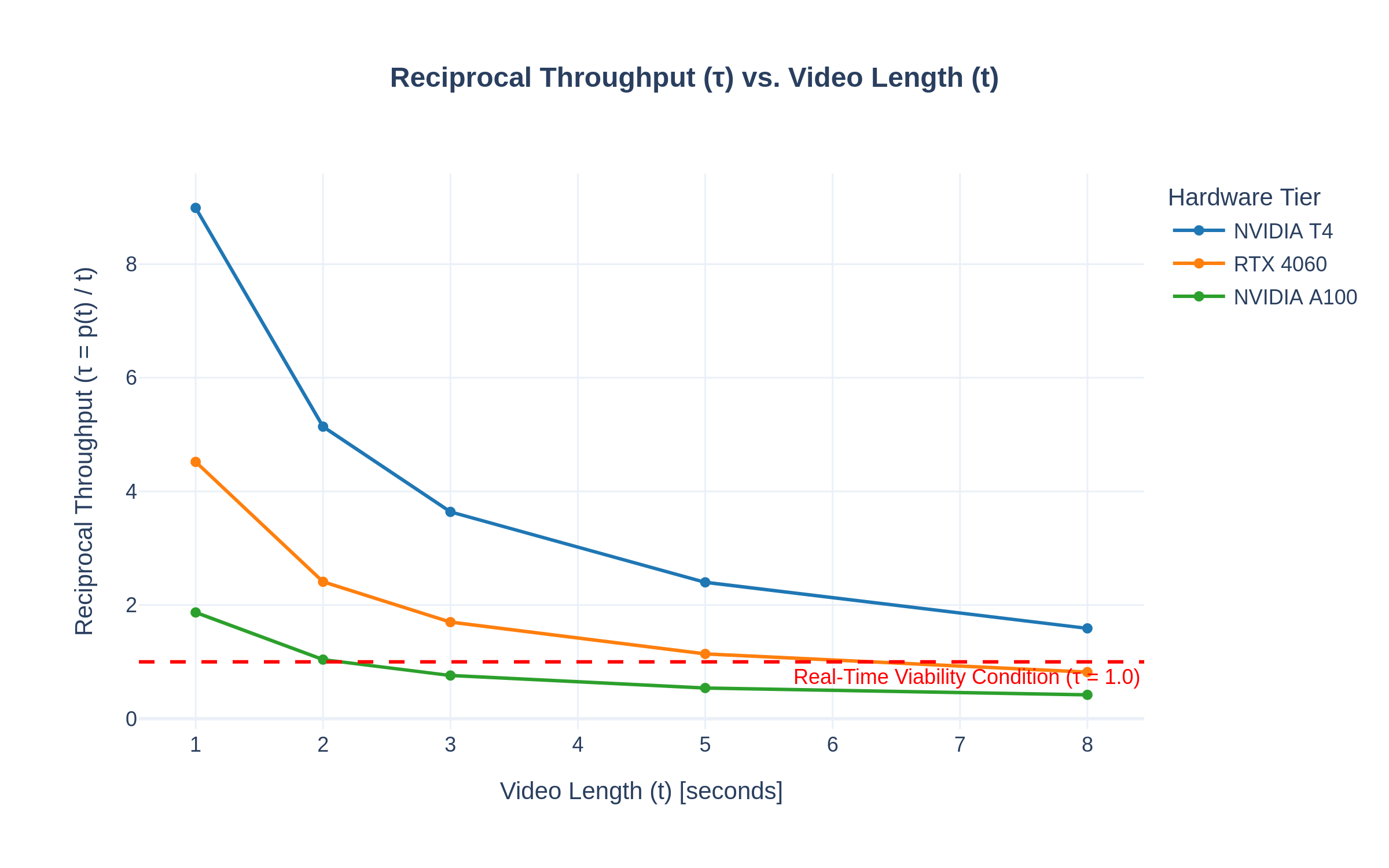}
    \caption{Reciprocal Throughput ($\tau(t)$) vs. Video Length ($t$) across the three hardware tiers. The horizontal dashed line at $\tau = 1.0$ represents the Real-Time Viability Condition, where processing time equals playback time. The system is stable only for $\tau < 1.0$.}
    \label{fig:throughput_vs_length}
\end{figure}

To better visualize these results in the context of our methodology, Figure~\ref{fig:throughput_vs_length} plots the Reciprocal Throughput ($\tau(t)$) as a function of video duration. This graph is the primary tool for validating our segmented processing protocol.

As predicted by our model (Equation~(\ref{eq:smooth})), Figure~\ref{fig:throughput_vs_length} provides direct empirical validation of our protocol. The enterprise-grade A100 GPU crosses the $\tau = 1.0$ threshold between $t=2$ and $t=3$ s, establishing an optimal segment duration ($T_{opt}$) of approximately 3 s for that hardware. The commodity-tier RTX 4060 achieves this condition between $t=5$ and $t=8$ s, making $T_{opt}$ around 8 s. In contrast, the baseline T4 GPU fails to achieve $\tau < 1.0$ within the tested 8-s range, operating entirely within the ``System Lag Regime.'' This directly confirms that the viability of the protocol is hardware-dependent and that smooth, real-time playback is readily achievable on modern commodity and enterprise~GPUs.

Furthermore, the data in Table~\ref{tab:objective_results} directly confirms the sub-linear scaling behavior that is foundational to our protocol. The disproportionately high processing time for short 1-s segments reveals the significant impact of fixed overheads from model loading and initialization. However, as video length increases, these initial costs are clearly amortized. For example, on the A100, doubling the video length from 1~s to 2~s increases the total processing time by only 11\% (from 1.87~s to 2.08~s), while the final 3 s of video (from 5~s to 8~s) add only 23\% to the total time. This flattening of the performance curve is precisely what enables our segmented protocol to work: by processing longer chunks, the system effectively ``catches up'' and can operate ahead of real-time.

\subsection{Subjective User Experience Analysis}\label{subsec:subjective}

Following the objective performance characterization, a subjective evaluation was conducted with 30 participants to assess the perceptual quality of the system's output. To test the core premise of our segmented processing protocol, participants were shown video clips generated under specific latency conditions corresponding to each hardware tier. 

For the baseline NVIDIA T4, which never achieved the real-time condition, users were shown the 8-s clip with its corresponding 12.7-s processing delay to gauge user tolerance for a lagging system. For the RTX 4060 and A100, participants were shown video clips with durations that satisfied the Real-Time Viability Condition ($\tau < 1.0$)—specifically, an 8-s clip for the RTX 4060 (6.6~s delay) and a 3-s clip for the A100 (2.3~s delay). This allowed for a direct evaluation of the ``Startup Delay Acceptability'' (SDA) under conditions where smooth playback is guaranteed.

Table~\ref{tab:subjective_results} summarizes the Mean Opinion Scores (MOS) for the five criteria. While the model outputs for LSA, MN, VIQ, and VOQ are technically identical across platforms, we report the collected scores for each condition to reflect any potential perceptual differences or biases introduced by the varied user experience.

\begin{table}[H]
    %\centering
    \caption{Mean Opinion Scores (MOS) from Subjective Evaluation (N~=~30). Scores are on a 5-point Likert scale. Presented as Mean $\pm$ Standard Deviation, with the 95\% Confidence Interval in brackets.}
    \label{tab:subjective_results}
    \small\setlength{\tabcolsep}{2.75mm}
    \begin{tabular}{lccccc}
        \toprule
        \textbf{Hardware} & \textbf{LSA} & \textbf{MN} & \textbf{VIQ} & \textbf{VOQ} & \textbf{SDA} \\
        \midrule
        \textbf{NVIDIA T4}  %MDPI: Please confirm if the bold formatting is necessary; if not, please remove it. The following highlights are the same.
        %AUTHOR: Confirmed. They are necessary.
        & 3.91 $\pm$ 0.88 & 3.76 $\pm$ 0.93 & 3.25 $\pm$ 1.12 & 4.55 $\pm$ 0.62 & 4.15 $\pm$ 0.79 \\
        \textit{(8~s video, 12.7~s delay)}  %MDPI: Please confirm if the italics are necessary; if not, please remove them. The following highlights are the same.
        %AUTHOR: Confirmed. They are necessary.
        & [3.58, 4.24] & [3.41, 4.11] & [2.83, 3.67] & [4.32, 4.78] & [3.85, 4.45] \\
        \midrule
        \textbf{RTX 4060} & 3.97 $\pm$ 0.85 & 3.81 $\pm$ 0.90 & 3.29 $\pm$ 1.12 & 4.58 $\pm$ 0.60 & 4.60 $\pm$ 0.61 \\
        \textit{(8~s video, 6.6~s delay)} & [3.64, 4.30] & [3.46, 4.16] & [2.87, 3.71] & [4.35, 4.81] & [4.38, 4.82] \\
        \midrule
        \textbf{NVIDIA A100} & 4.02 $\pm$ 0.83 & 3.86 $\pm$ 0.89 & 3.33 $\pm$ 1.10 & 4.62 $\pm$ 0.58 & 4.85 $\pm$ 0.42 \\
        \textit{(3~s video, 2.3~s delay)} & [3.70, 4.34] & [3.52, 4.20] & [2.91, 3.75] & [4.40, 4.84] & [4.70, 5.00] \\
        \bottomrule
    \end{tabular}
\end{table}

\begin{figure}[H]
  %  \centering
    % Replace the \rule command with your exported figure file
    \includegraphics[width=0.8\textwidth]{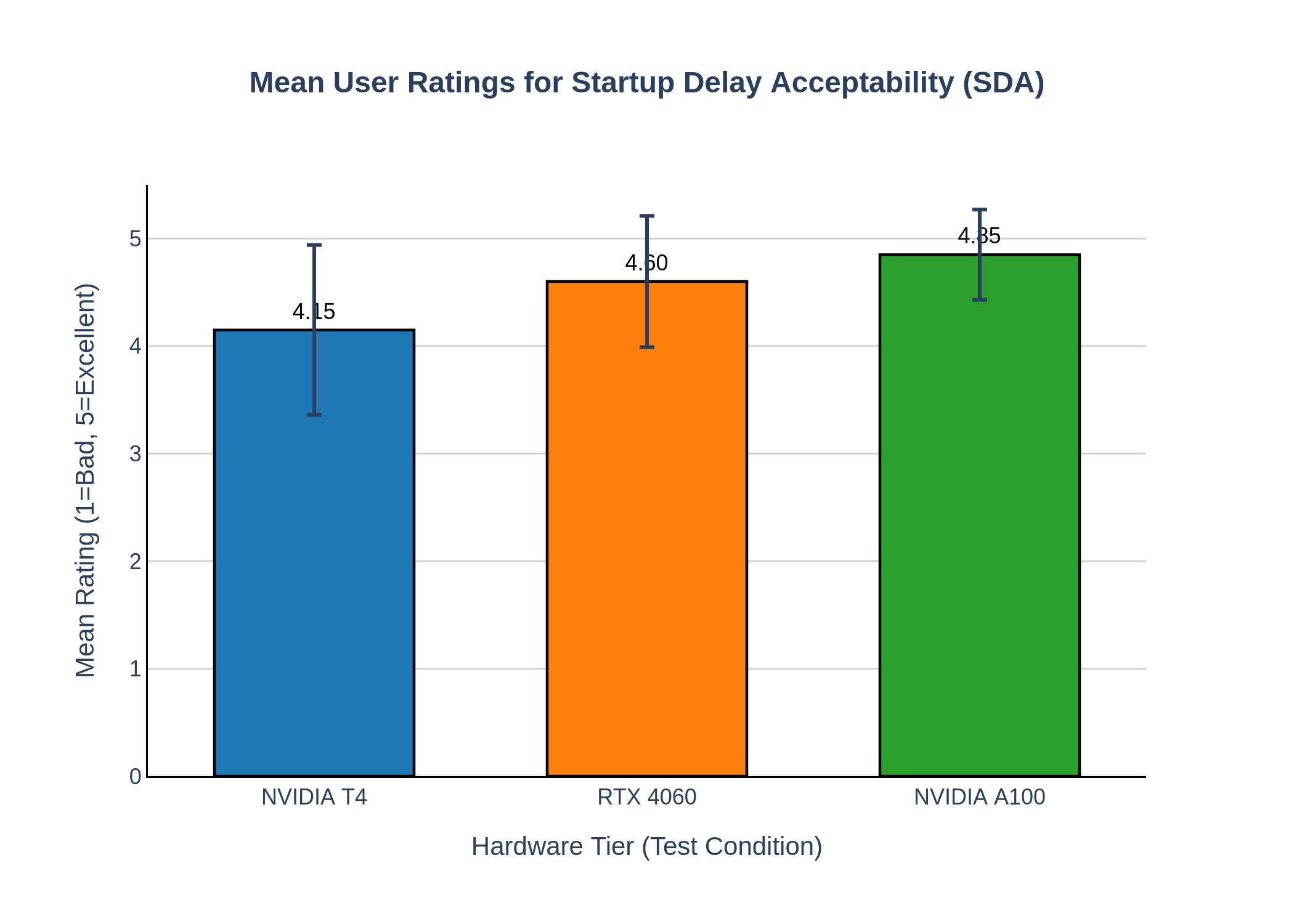}
    \caption{Bar chart of Mean User Ratings (N~=~30) for Startup Delay Acceptability (SDA). The height of each bar represents the mean score for a given hardware condition, and the error bars indicate the standard deviation of the user ratings.}  %MDPI: Please confirm whether the overlapping in this figure affects scientific understanding and if it does, please revise it.
    % %AUTHOR: Since the values of the figure are mentioned within the text, we beleive it does not affect scientific understanding.
    \label{fig:bar_chart}
\end{figure}

To provide a clear visualization of the user ratings for the most critical perceptual metric, Figure~\ref{fig:bar_chart} presents a bar chart of the mean scores for Startup Delay Acceptability~(SDA).

The analysis reveals two key insights. First, the perceptual quality of the generative models themselves was consistent, regardless of the underlying hardware. The high ratings for Vocal Quality (VOQ), with a mean score consistently above 4.5, and positive results for Lip Sync Accuracy (LSA) and Motion Naturalness (MN) confirm the effectiveness of the chosen models. As indicated by the MOS scores, Visual Quality (VIQ) remained the lowest-rated metric across all tests, signifying that visual artifacts from the LipSync model are the primary bottleneck in perceptual quality.

Second, and most critically, the bar chart in Figure~\ref{fig:bar_chart} provides direct user validation for our segmented processing protocol by illustrating its hardware-dependent nature. There is a clear and strong positive correlation between hardware performance and user satisfaction with the startup delay. The baseline T4, with a long 12.7~s delay, received a good but comparatively low mean SDA score of 4.15. This score significantly increased to 4.60 for the RTX 4060, which halved the delay to 6.6~s. Finally, the A100, which reduced the startup delay to a near-imperceptible 2.3~s, received an almost perfect mean score of 4.85.

This is a key finding. It provides direct empirical validation for the core trade-off in our protocol: users find the initial, predictable startup delay to be an acceptable price for a subsequent smooth, uninterrupted playback experience. Furthermore, it proves that as hardware capabilities improve, the perceived trade-off diminishes to the point of being a non-issue, confirming that our protocol is not only technically sound but creates a highly satisfactory and user-centric Quality of Experience on modern hardware.

% ===============================================================
% ===============================================================
% ===============================================================

\section{Discussion}\label{sec:discussion}

The results presented in the previous section provide strong empirical validation for our proposed system-level framework. This section moves beyond the raw data to interpret these findings, discuss their broader implications for the field of real-time generative AI systems, and honestly assess the limitations of the current work to guide future research.

\subsection{Empirical Validation of the System Architecture \& Protocols}

Our primary contribution is an architectural and protocol-level solution to the challenges of latency and scalability in multi-user generative AI applications. The objective performance analysis provides three key points of validation for this framework.

First, the sub-linear scaling of inference time ($p(t)$), as evidenced by the data in Table~\ref{tab:objective_results}, is not merely a performance artifact but a fundamental property that our segmented processing protocol is designed to exploit. The initial high overheads for short-duration clips confirm that a naive, continuous streaming approach would be highly inefficient and prone to failure. By demonstrating that these overheads are amortized over longer segments, we have empirically validated the core assumption of our protocol: processing in optimized, fixed-length chunks is demonstrably more efficient.

Second, the multi-tiered hardware evaluation (Figure~\ref{fig:throughput_vs_length}) confirms that our ``Real-Time Viability Condition'' ($\tau(t) < 1.0$) is a practical and achievable target. The results draw a clear and actionable distinction between hardware tiers: while the system is \textit{feasible}  %MDPI: Please confirm if the italics are necessary; if not, please remove them. The following highlights are the same.
%AUTHOR: Confirmed. They are necessary.
on a baseline cloud GPU like the NVIDIA T4, it becomes \textit{production-ready} on both modern commodity hardware (RTX 4060) and enterprise-grade infrastructure (NVIDIA A100). This finding is significant as it provides a clear roadmap for deployment: developers can prototype on accessible hardware, confident that the system will achieve true real-time performance when scaled to production environments, thereby de-risking the adoption of such technologies.

Third, the strong positive rating for Startup Delay Acceptability (SDA) in our subjective evaluation provides crucial user-centric validation for our engineering trade-off. The data confirms that users perceive a system with a predictable, one-time initial latency followed by smooth, uninterrupted playback as a high-quality experience. This psycho-visual finding challenges the conventional wisdom that all latency must be minimized at all costs. Instead, it suggests that for computationally intensive tasks, predictability and reliability of the stream are more critical to the user's Quality of Experience than the absolute initial delay. This insight has broad applicability beyond video translation to other real-time generative tasks.

{Furthermore, our multi-tiered subjective evaluation can be interpreted as an implicit ablation study of our protocol's primary benefit. The NVIDIA T4 test case, which fails to meet the real-time viability condition ($\tau > 1.0$), represents a ``control'' system where users experience the full, raw processing delay without the guarantee of smooth playback. The RTX 4060 and A100 test cases, which satisfy the condition ($\tau < 1.0$), represent the ``treatment'' system operating as designed. The statistically significant and substantial increase in the Startup Delay Acceptability (SDA) score from the control to the treatment conditions provides direct quantitative evidence of our protocol's effectiveness in improving the user's Quality of Experience.}

\subsection{Analysis of User Perception and Pipeline Quality}

The subjective evaluation, now fortified with a larger participant pool, allows for a more confident analysis of the end-user experience. The standout result is the exceptionally high rating for Vocal Quality (VOQ), which, with mean scores consistently above 4.5, indicates that modern zero-shot voice cloning technology is mature enough for practical applications. This success is critical, as preserving the speaker's vocal identity is paramount for maintaining conversational presence and authenticity.

Conversely, the lowest-scoring metric, Visual Quality (VIQ), points to a clear area for future improvement. The wider distribution of scores for VIQ suggests that while some users were not bothered by visual artifacts, a significant portion found them distracting. This indicates that the chosen LipSync model (Wav2Lip-GAN), while functional, represents the primary bottleneck in perceptual quality. This finding underscores the modularity of our proposed pipeline; the system architecture itself is robust, and as more advanced, diffusion-based or NeRF-based lip-sync models become available, they can be swapped in to directly address this weakness and improve the overall user experience.\

\subsection{Limitations and Future Work}

While this work successfully demonstrates a viable system-level framework, it is essential to acknowledge its limitations to provide a clear path for future research.

First, our performance analysis deliberately isolates computational latency from network latency to provide a clean characterization of the system's processing capabilities. A real-world deployment would need to integrate this architecture with robust network protocols capable of handling jitter, packet loss, and variable bandwidth to ensure a seamless end-to-end experience. Our work provides the computational foundation upon which such a system can be built.

Second, the ``Token Ring'' mechanism is presented at an architectural level. This study does not prescribe a specific implementation for the token-passing logic (e.g., manual ``raise hand'' features, automatic voice activity detection, or moderated control). The development and evaluation of these different token management strategies represent a rich area for future work at the intersection of system design and Human-Computer Interaction (HCI). The critical contribution of our work is the formal demonstration that \textit{any} such turn-taking mechanism, once implemented, fundamentally resolves the quadratic scalability problem.

{Finally, while our evaluation provides a thorough validation of our framework's internal performance claims and user acceptability, this study does not include a direct, quantitative comparison against other end-to-end system architectures. Our assessment was designed to prove the effectiveness of our specific architectural contributions using system-level metrics such as Reciprocal Throughput and Startup Delay Acceptability. As the field matures, the establishment of standardized, system-level benchmarking protocols will be a crucial next step, enabling researchers to directly compare the performance, scalability, and efficiency of different architectural approaches.}

% ===============================================
% ===============================================
% ===============================================

\section{Conclusions}\label{sec:conclusion}

The integration of generative AI into real-time communication systems offers the transformative potential to dissolve language barriers, yet it is hindered by formidable system-level challenges of latency and scalability. This paper has addressed these challenges by introducing a novel, comprehensive framework for the deployment of real-time multilingual video translation. Our primary contributions are twofold: a scalable \textit{Token Ring} system architecture %MDPI: Please confirm if the bold formatting is necessary; if not, please remove it. The following highlights are the same.
%AUTHOR: Confirmed. 
 that reduces the computational complexity of multi-user meetings from an intractable $\mathcal{O}(N^2)$ to a manageable $\mathcal{O}(N)$, and a \textit{Segmented Batched Processing} protocol with inverse throughput thresholding, designed to manage the high intrinsic latency of generative pipelines.

Through the development of a proof-of-concept and a rigorous, multi-tiered empirical evaluation, we have validated the efficacy of this framework. Our objective analysis experimentally demonstrated that the system achieves the critical \textit{Real-Time Viability Condition ($\tau < 1.0$)} on both modern commodity and enterprise-grade hardware, confirming its feasibility for widespread deployment. Furthermore, our statistically robust subjective evaluation revealed that users find the system's trade-off of a predictable initial latency for smooth, uninterrupted playback to be highly acceptable, resulting in a positive overall user~experience.

While limitations regarding network integration and visual quality remain areas for future model-level and implementation-level research, this work provides a foundational, validated system architecture. By solving the critical bottlenecks of scalability and latency management, this framework offers a practical roadmap for developers and researchers to build the next generation of inclusive, effective, and truly global communication platforms.

{Future work will focus on two critical frontiers. First, the integration of more advanced, non-autoregressive generative models for the pipeline's bottleneck—the LipSync stage. Emerging diffusion-based and NeRF-based talking head models, while computationally demanding, offer superior visual fidelity and could be integrated into our framework to specifically address the limitations in Visual Quality (VIQ) identified in our user study. Our architecture provides the ideal testbed for quantifying the system-level latency trade-offs of these next-generation models. Second, we will explore dynamic, adaptive segmentation protocols. While our current protocol uses a fixed optimal segment duration ($T_{opt}$), a more sophisticated approach could dynamically adjust the chunk size in real-time based on network conditions and the linguistic complexity of the source content, further optimizing the balance between latency and computational throughput for a truly seamless user~experience.}

%%%%%%%%%%%%%%%%%%%%%%%%%%%%%%%%%%%%%%%%%%
\vspace{6pt}

\section*{Ethical statement}
Informed consent was obtained from all subjects involved in the study. Generative AI tools were used solely to improve language clarity. The authors take full responsibility for the content.

\section*{Data availability}
The authors agree to share datasets upon requests from readers.

\section*{Conflicts of interest}
The authors declare no conflict of interest.

\section*{References}
%\bibliography{refs}

%%%%%%%%%%%%%%%%%%%%%%%%%%%%%%%%%%%%%%%%%%
%% for journal Sci
%\reviewreports{\\
%Reviewer 1 comments and authors’ response\\
%Reviewer 2 comments and authors’ response\\
%Reviewer 3 comments and authors’ response
%}
%%%%%%%%%%%%%%%%%%%%%%%%%%%%%%%%%%%%%%%%%%
\end{document}